\documentclass[]{aa}
\usepackage[varg]{txfonts}
\usepackage{graphicx}                                                        
\usepackage{graphics} 
\usepackage{epsfig}
\usepackage{mathptmx} 
\usepackage{times} 
\usepackage{amssymb} 
\usepackage{array}
\usepackage{hyperref}
\usepackage{caption}
\usepackage{natbib}
\usepackage{sidecap}
\usepackage{longtable}
\usepackage[hang]{footmisc} 
\begin{document}
\newcommand{\NAT}[2][]{}

\title{Long-term variations in the X-ray activity of HR~1099}

\author{V. Perdelwitz\inst{1}
  \and F.H. Navarrete\inst{2}
  \and J. Zamponi\inst{2}
  \and R.E. Mennickent\inst{2}
  \and M. V\"olschow\inst{1}
  \and J. Robrade\inst{1}
  \and P.C. Schneider\inst{1}
  \and D.R.G. Schleicher\inst{2}
  \and J.H.M.M. Schmitt\inst{1}
  }

\institute{Hamburger Sternwarte, Universit\"at Hamburg, Gojenbergsweg 112, 21029 Hamburg, Germany
  \and Departamento de Astronom\'ia, Facultad Ciencias F\'isicas y Matem\'aticas, Universidad de Concepci\'on, Av.  Esteban Iturra s/n Barrio Universitario, Casilla 160-C, Concepci\'on, Chile}

 \date{Received: 1 November 2017 / Accepted: 5 June 2018}

\abstract 
{Although timing variations in close binary systems have been studied for
a long time, their underlying causes are still unclear. A possible
explanation is the so-called Applegate mechanism, where a strong,
variable magnetic field can periodically change the gravitational
quadrupole moment of a stellar component, thus causing observable period
changes. One of the systems exhibiting such strong orbital variations is
the RS CVn binary HR~1099, whose activity cycle has been studied by
various authors via photospheric and chromospheric activity indicators,
resulting in contradicting periods.}
{We aim at independently determining the magnetic activity cycle of
HR~1099 using archival X-ray data to allow for a
comparison to orbital period variations.}
{Archival X-ray data from 80 different observations of HR~1099 acquired 
with 12
different X-ray facilities and covering almost four decades were used to
determine X-ray fluxes in the energy range of $2-10$~keV via spectral
fitting and flux conversion. Via the Lomb-Scargle periodogram
we analyze the resulting long-term X-ray light curve to search for
periodicities.}
{We do not detect any statistically significant periodicities within 
the X-ray data.
An analysis of optical data of HR~1099 shows that the derivation of
such periods is strongly dependent on the time coverage of available data,
since the observed optical variations strongly deviate from a pure sine 
wave.
We argue that this offers an explanation as to why other authors derive 
such a wide
range of activity cycle periods based on optical data.
We furthermore show that X-ray and optical variations are correlated in 
the sense that the star tends to be optically fainter when it is X-ray 
bright.}
{We conclude that our analysis constitutes, to our knowledge, the 
longest
stellar X-ray activity light curve acquired to date, yet the still rather
sparse sampling of the X-ray data, along with
stochastic flaring activity, does not allow for the independent
determination of an X-ray activity cycle. }


\maketitle

 \section{Introduction}
Stellar activity cycles, first discovered for our Sun by \cite{1844AN.....21..233S}, have been observed in various stars via photospheric and chromospheric activity indicators, both in single and binary systems \citep[e.g.,][]{2005LRSP....2....8B,2007AJ....133..862H}. With the advent of space-borne X-ray observatories it has become possible to extend this research to the X-ray range, which has been shown to be a robust tracer of activity cycles \citep{2004A&A...418L..13F,2012A&A...543A..84R,2013A&A...559A.119L,2017MNRAS.464.3281W}.\\
In addition to enabling the study of the underlying dynamo, magnetic activity cycles have been suggested to be the cause of timing variations in close binary systems.
These have been observed over a broad range of such systems, including Algol and RS~CVn systems, with relative period modulation amplitudes of  $\Delta P/P\sim10^{-5}$ and typical cycle periods of $30-50$~yrs, as well as cataclysmic variables and W~UMa systems with $\Delta P/P\sim10^{-6}$ and cycles of $5-30$~yrs \citep[e.g.,][]{Warner88, Hall89}. Post-common-envelope binaries (PCEBs) exhibit a very similar behavior, with $\Delta P/P\sim10^{-6}$ and modulation periods of $10-30$~yrs \citep{Zorotovic13, Bours16}.\\
\noindent 
A simple explanation of these period variations is to invoke the presence of a third body, giving rise to the variation via the Light Travel Effect (LTE) effect (see e.g., \cite{2012AN....333..754P}). 
However, while the existence of such circumbinary companions has been postulated to explain timing variations in various systems, to date none of them has been confirmed with an independent method; for example, the third body causing timing variations in V471~Tau first proposed by \cite{1986ApJ...300..785B} could not be confirmed with a direct imaging attempt with VLT-SPHERE \citep{Hardy15}.\\ 
\noindent 
It is therefore essential to assess whether other mechanisms can produce the observed 
period variations. Such a mechanism could be a powerful magnetic activity cycle as expected in the presence of rapidly rotating and convective stars \citep{Baliunas96}.  During the course
of a cycle, the stellar angular momentum could be redistributed, thus 
leading to quasi-periodic changes of the stellar quadrupole moment and orbital period. 
This model, originally proposed by \citet{Applegate92}, has been improved by \citet{Lanza99} by adopting a consistent formulation for virial equilibrium. \citet{Brinkworth06} extended the model by introducing a finite-shell formalism, showing that the latter substantially increases the energy required to drive the Applegate mechanism. \citet{Volschow16} applied this model to a total of $16$ close binary systems and showed that the mechanism is clearly feasible in four of the systems, while it can be ruled out on energetic grounds in eight cases; an additional amount of four systems remained unclear. 
Further, improved models for the Applegate mechanism have been put forward by \citet{Lanza05, Lanza06a}. The authors found that if the observed period variations were driven by the Applegate mechanism, the energy dissipation in the convection zone would exceed the
star's luminosity. As shown recently, the resulting changes of the quadrupole moment may also affect the mass transfer rate between the stars, and potentially give rise to the long cycles in Double Periodic Variables \citep{2017A&A...602A.109S}.\\
\noindent 
One of the systems exhibiting substantial period variations is HR~1099 (or V711 Tau), 
a RS CVn binary of spectral type K1~IV+G5~V with an orbital period of 2.84~days first 
identified by \cite{1976AJ.....81..771B}. Its proximity of 30.7~pc \citep{2007A&A...474..653V} 
and its very high level of magnetic activity in all available indicators have made it one of the 
prime candidates for the study of stellar dynamos (see e.g., \cite{1999MNRAS.302..457D}, \cite{GarcaAlvarez2002}, \cite{2004ApJS..153..317O} and \cite{2015MNRAS.449.1380C}).
Several authors have searched for magnetic activity periods in HR~1099 via long-term photometry, which is sensitive to detecting
variations in the spot coverage.   Such activity cycles have been reported by  \citet{2006A&A...455..595L}, who obtain a $19.5\pm2$~yr period, by \citet{2007ApJ...659L.157B}, who derive a period of $15-16$~yr,  and by \citet{2010A&A...521A..36M} who
report a $14.1\pm0.3$~yr period.   An inspection of the light curves shows clear periodicities despite a very large
scatter in the optical photometry.
In addition to their activity period derivation, \citet{2010A&A...521A..36M} show that the O-C diagram of HR~1099 can be improved considerably by fitting the orbital period of the system to the O-C data, resulting in a well-defined sinusoidal shape with a period of $36.3\pm1.9$~yrs.\\
\noindent 
As a consequence, the question arises whether magnetic activity is also periodic in other proxy indicators. \cite{2014ApJ...783....2D} have carried out study into the long-term X-ray variability of AR Lac found that it is constant and displays no signs of a coronal activity cycle.
In this paper 
we therefore independently investigate the magnetic activity period of HR~1099 using an X-ray light curve spanning almost 40~yrs in order to test whether this method is a valid means to test the Applegate mechanism. 
A theoretical motivation is given in Section~\ref{theory} and the observational data are described in Section~\ref{observations}. A summary and discussion of our results is given in Section~\ref{discussion}.

\section{Theoretical motivation}\label{theory}
 We show in the following that a long-term magnetic activity cycle can be qualitatively understood through simple dynamo models. For this purpose, we employ the relation between rotation period and activity cycle given as \citep{Soon93, Baliunas96}
\begin{equation}\label{eq:cyc-rot}
 P_{cycle} = D^\alpha P_{rotation},
\end{equation}
where $D$ is the dynamo number and $\alpha$ is a parameter of order $0.5$. Furthermore, $D$ is related to the Rossby number, $Ro$, via 
\begin{equation}\label{eq:D}
 D = Ro^{-2}
\end{equation}
with Ro $\propto \text{P}/\tau$, where P is the period and $\tau$ is the convective turnover time, can also be obtained from \citep{Soker00}
\begin{equation}\label{eq:ro}
 Ro = 9\left( \frac{v_c}{10\,\textnormal{km s}^{-1}} \right)\left( \frac{H_p}{40\,R_\odot} \right)^{-1}\left( \frac{\omega}{0.1\,\omega_{Kep}} \right)^{-1}\left( \frac{P_{Kep}}{1\,\textnormal{yr}} \right),
\end{equation}
with $v_c$ denoting the convective velocity, $H_p$ the pressure scale height, and $\omega$, $\omega_{Kep}$, and $P_{Kep}$ the angular velocity,  Keplerian angular velocity, and Keplerian orbital period, respectively.
\begin{figure}[ht!]
\centering
\includegraphics[width=0.4\textwidth]{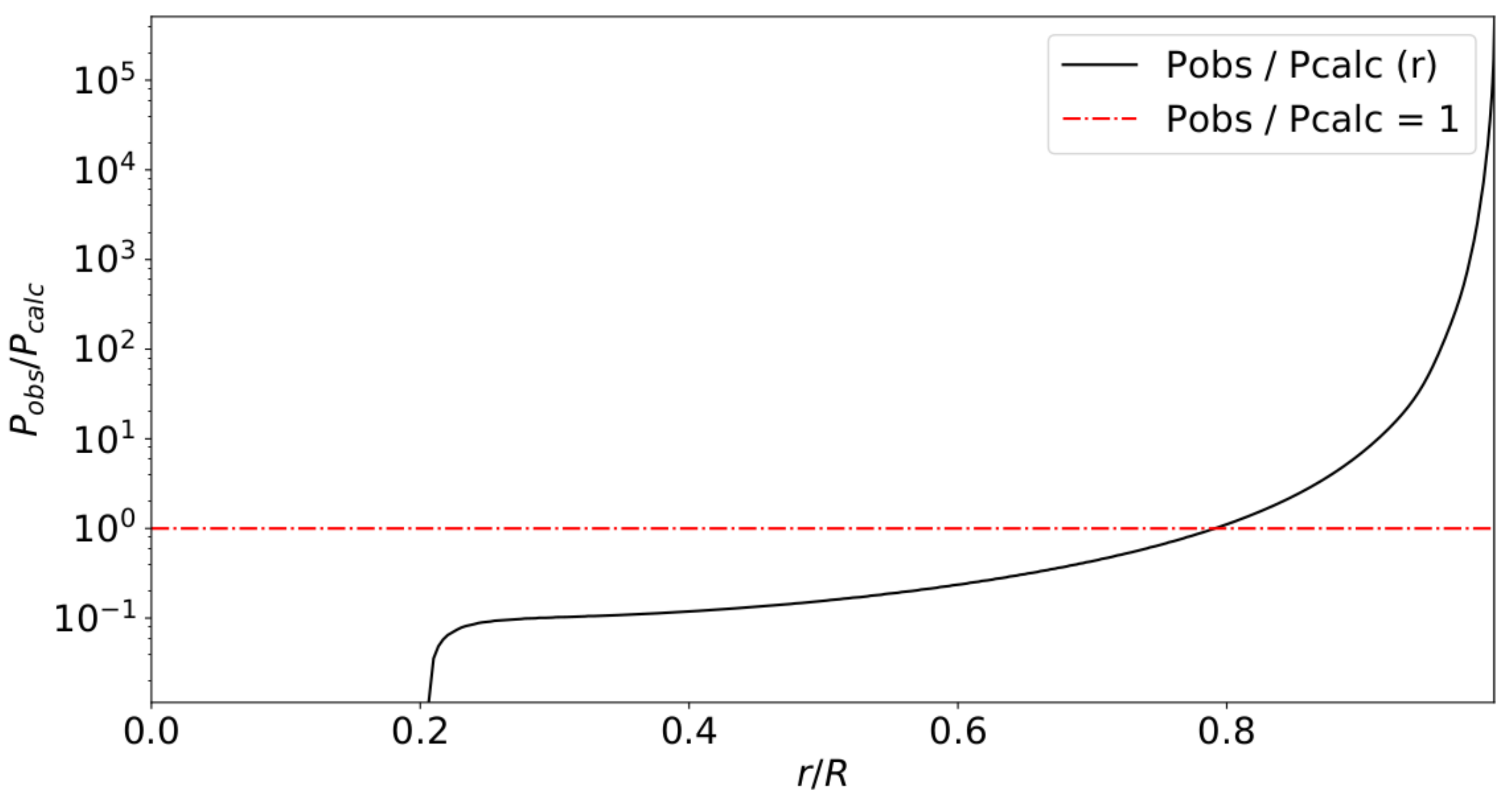}
\captionsetup{font=footnotesize}
\caption[width=0.45\textwidth]{Radial profile of $P_{obs} / P_{calc}$ obtained from Eq.~\ref{eq:cyc-rot} (black curve) and $P_{obs} / P_{calc} = 1$ (horizontal red line). The lines meet at $r/R \sim 0.8$, suggesting that the observed activity cycle is driven in this region.}
\label{radperiod}
\end{figure}
\noindent The convective velocity and the pressure scale height are properties inside the star, which we calculate using the  stellar evolution code MESA \citep{2011ApJS..192....3P}. We first evolve a Sun-like star with solar-like rotation up to $4.7\,\textnormal{Gyr}$, which we use to calibrate $\alpha$ in Eq.~\ref{eq:cyc-rot}. The model employs a solar metallicity, and we choose mixing length theory \texttt{MLT = 'ML1'} with $\alpha_{MLT} = 1.5$. Considering that the solar dynamo is expected to take place at the bottom of the convection zone with $R\sim 0.7\,R_\odot$, we use the properties at that point and find that $\alpha \sim 0.86$.
\noindent We subsequently run a simulation for the subgiant in HR~1099 assuming that it is tidally locked, so that its rotation corresponds to a fixed fraction of the critical angular velocity with $\Omega_{ZAMS} / \Omega_{crit} = 0.28$ following from the binary rotation period\footnote{Here, we use the the notation of \cite{2013ApJS..208....4P}}. The mass of the subgiant is set to $M=1.3\,M_\odot$ based on \cite{1976AJ.....81..771B}. The run is stopped when the radius is $R=4.0\,R_\odot$.
\noindent We calculate the dynamo number $D$ as a function of radius from the convection velocity $v_c$ and the pressure scale height $H_p$, allowing us to compute the expected dynamo activity cycle as a function of radius. In Figure~\ref{radperiod}, the black curve shows the fraction $P_{obs} / P_{calc}$ as a function of the radial coordinate $r/R$ and the horizontal red line is $P_{obs}/P_{calc} = 1$, where $P_{obs}$ is assumed here to be of the order of 20 yrs. They intersect at $r/R \sim 0.8$. \\
While based on simplifying assumptions, the result suggests that a dynamo cycle of $\sim20$~yrs would be predominantly driven in the outer parts of the shell. 
\noindent We further make a preliminary assessment if such a dynamo cycle may be responsible for the O-C variations observed in the system through the Applegate mechanism. As a remark, we note that the ratio between activity cycle and O-C variation cycle is approximately 1:2, as is the case in previously observed systems \citep[e.g.,][]{Lanza01}. 
The expected ratio is however not well explored in theoretical models. To assess whether the Applegate mechanism is energetically feasible, we employ the formalism developed by \citet{Volschow16} using the two-zone model, which has been implemented in a publicly available \texttt{Applegate calculator}\footnote{\url{http://theory-starformation-group.cl/applegate/}}. 
In this calculation, we neglect the mass contribution from the radiative core, where dynamo effects and related changes of the angular momentum cannot be expected to be relevant \citep[see also discussion in][]{Lanza06a}. We first recalculate the $k_1$ and $k_2$ coefficients defined by \citet{Volschow16} based on the MESA output neglecting the central core. The lowest energy is found when we consider the exchange of angular momentum between the shells with $0.2 < r/R_{in} < 0.65$ and $0.65 < r/R_{out} < 1$. 
The result suggests that the Applegate mechanism may be energetically feasible, and would be driven somewhat more in the interior of the star.

\section{Observations}\label{observations}
\label{hr1099}
HR~1099 is one of the brightest known coronal X-ray sources. 
Therefore HR~1099 has been observed -- starting in 1978 -- by virtually all X-ray facilities, 
even those with small effective areas and lower sensitivity, thus forty years of X-ray observations are available.  
Specifically, we have identified a total of 78 data sets, 60 in the form of spectra and 18 as count rates,
from 12 different facilities. 
Many of these observations have been carried out for calibration purposes or with the goal of studying stellar flares \citep{1999ApJ...524..988K,2001A&A...365L.324B,2004ApJS..153..317O,2016PASJ...68...90T}, yet, to our knowledge, no long-term activity study of HR~1099 based on several data sets has been carried out so far.
In Tab.~\ref{tab:obs} we provide a detailed list of these X-ray observations with the appropriate references.

 \begin{table}
\centering
\begin{tabular}{llccc} 
\hline\hline 
Mission & Instrument & energy range &ref.&\#\\
& & [keV] &&\\ 
\hline \\
HEAO1& A2 & 2 - 60 &[1]& 5\\ 
Einstein& IPC & 0.4 - 4 &[2]& 4\\ 
 & SSS & 0.5 - 4.5 && 4\\ 
EXOSAT&  ME & 1 - 50&[3] & 8\\ 
ROSAT& PSPC & 0.08 - 2.9 &[4]& 2\\ 
 & HRI & 0.1 - 2.4 && 9\\ 
 Ginga& LAC & 1.5 - 37 &[5]& 3\\ 
ASCA& SIS0,1 & 0.3 - 12 &[6]& 2\\ 
 & GIS2,3 & 0.4 - 12 && 2\\ 
RXTE& PCA & 1.4 - 100 &[7]& 17\\ 
BeppoSAX& LECS & 0.1 - 14 &[8]& 1\\ 
 & MECS & 0.8 - 12 && 1\\ 
XMM-Newton& EPIC PN & 0.1 - 12 &[9]& 8\\ 
 & MOS1& 0.1 - 12 && 1\\ 
Chandra& HETG,ACIS-S & 0.4 - 8 &[10]& 2\\ 
Swift& XRT& 0.2 - 10 &[11]& 9\\ 
MAXI& GSC & 2 - 20 &[12]& 4\\
 & SSC & 0.7 - 7 && 4\\ 
\hline
\end{tabular}
\captionsetup{width=\linewidth}
\caption[width=\textwidth]{Instruments in use. The last column shows the total number of spectra analyzed in this work. For some instruments spectra from multi-instrument missions were fitted simultaneously (see Section~\ref{sec:sd}).  Instrument references: \cite{1979SSI.....4..269R}[1]; \cite{1979ApJ...230..540G}[2]; \cite{1981SSRv...30..479T}[3]; \cite{1982AdSpR...2..241T}[4]; \cite{1989PASJ...41..345T}[5];  \cite{1994PASJ...46L..37T}[6]; \cite{1996SPIE.2808...59J}[7]; \cite{1997A&AS..122..299B}[8]; \cite{2001A&A...365L...1J}[9]; \cite{2002PASP..114....1W,2005PASP..117.1144C}[10]; \cite{2004ApJ...611.1005G}[11]; \cite{2009PASJ...61..999M}[12].}
\label{tab:obs}
\end{table}

\begin{figure*}
\centering
\includegraphics[width=17cm]{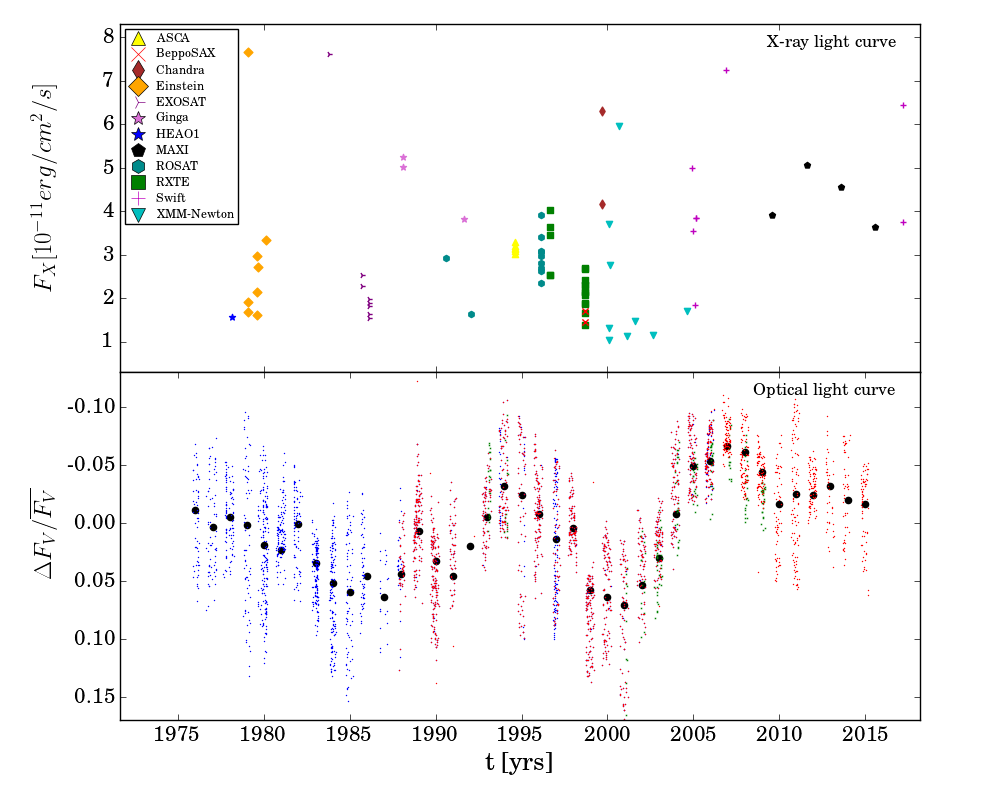}
%\captionsetup{width=0.75\textwidth,font=footnotesize}
\caption{Light curves of HR 1099 in the X-ray and optical regime. Top panel: X-ray flux of HR~1099 in the energy range 2-10~keV as a function of time. For clarity, we omitted the error bars in this plot, but the values can be found in Table~\ref{tab:data} . Bottom panel: Change in optical flux (V-band) as a function of time.
Blue data are taken from \cite{2007ApJ...659L.157B}, green data from \cite{2010A&A...521A..36M} and red data from \cite{2017ApJ...838..122J}. The black dots represent an annual binning of the data.}
\label{xraylcunfolded}

\end{figure*}

For the derivation we have to distinguish between two types of data, namely (i) spectra and (ii) data without spectral information, which we converted into fluxes assuming a model based on XMM-Newton spectra.

\subsection{Spectral data}
\label{sec:sd}
The majority of the spectral X-ray data of HR~1099 are available as spectral pipeline products from the 
NASA HEASARC\footnote{\url heasarc.gsfc.nasa.gov} data base; we used the HEASARC data whenever available. Specifically, we custom processed the XMM-Newton data (acquired from the ESA archive) with the Science Analysis System \citep{2004ASPC..314..759G} and downloaded Chandra spectra from the TGCat database \citep{2011AJ....141..129H}. MAXI data were obtained via the on demand tool on the project website\footnote{\url maxi.riken.jp}, with a binning of 2~yrs due to the small effective area of the instrument and target and background apertures of 2 and 3~deg. SWIFT data were generated with the Build Swift-XRT products tool \citep{2009MNRAS.397.1177E} on the SWIFT homepage\footnote{\url swift.ac.uk}.\\ 
For the following spectral analysis we used XSPEC version 12.7.1 \citep{1996ASPC..101...17A}. The 60 spectra were grouped with the grppha task to a minimum of 15~cts per spectral bin and clipped to the range of 2-10~keV, a spectral range covered by all but one instrument, the Einstein SSS. Due to an erroneous normalization of the pipeline products of the Einstein SSS and MPC, it was not possible to fit spectra from both instruments simultaneously and thereby cover the entire wavelength range. This and the presence of an instrumental artifact at $\approx 3.5$~keV led us to utilize only the SSS data and extrapolate the spectra beyond 3~keV by using the XSPEC energies and adding 100 additional linear bins up to 10~keV.\\
The first spectral fit was performed with a cflux*(mekal+mekal) model, where the model abundances were linked to each other, leaving  six free parameters: flux, kT$_1$, norm$_1$, kT$_2$, norm$_2$ and abundance. All spectra with a reduced $\chi^2>2$ were then re-fitted after adding a third mekal component to the model. We were merely interested in obtaining a good flux estimate in the energy range in question, and not in any coronal parameters such as plasma temperatures or abundances. For the majority of the spectra our fits yielded a reduced $\chi^2$ in the range of 0.63-1.8 (see Tab.~\ref{tab:data}), the exceptions being one Einstein SSS and two RXTE spectra.\\ 
The output fluxes from the XSPEC cflux model were used for further analysis along with flux uncertainties obtained with the XSPEC error function for a $90\%$ confidence level.\\
In the case of multi-instrument missions with more than one instrument covering 2-10~keV, all spectra were fitted simultaneously, with the exception of the photon-starved BeppoSAX data, where a simultaneous fit yielded a reduced $\chi^2>3$.

\subsection{Flux conversion}
We converted the 18 data sets from those instruments, for which only count rates were available (ROSAT, Einstein IPC, Ginga) into physical fluxes by computing conversion factors in the following manner. 
Two XMM-Newton spectra were fitted with a 3-T apec model in the spectral range 0.2-12.0~keV, one with a medium flux level as determined in the analysis described in the previous Section (Obs.-ID 0116890901), and one with the lowest flux of the entire XMM-Newton data set (Obs.-ID 0116200701). After the first fit, the plasma temperatures and abundances were frozen to the closest value available in the HEASARC WebPIMMS tool, after which the fit is repeated.
The determined plasma temperatures and relative emission measures, along with the model component normalizations, were used as input for WebPIMMS, and a conversion factor was determined for all instruments in question. The conversion factors derived via the XMM-Newton Obs.-ID 0116890901 were then used to convert instrument count rates into physical fluxes, while the difference between the mid- and low-state conversion factors was used as a systematic error. The inclusion of this systematic error was necessary due to the fact that a flux conversion without any a priori knowledge of the coronal properties during a given observation must consider the entire range of system parameters representative of the system. All conversion factors are displayed in Tab.~\ref{tab:data} along with the other data.

\subsection{Light curve analysis}
\label{pd} 
\subsubsection{Period analysis}
\begin{figure}
\centering
\includegraphics[width=0.5\textwidth]{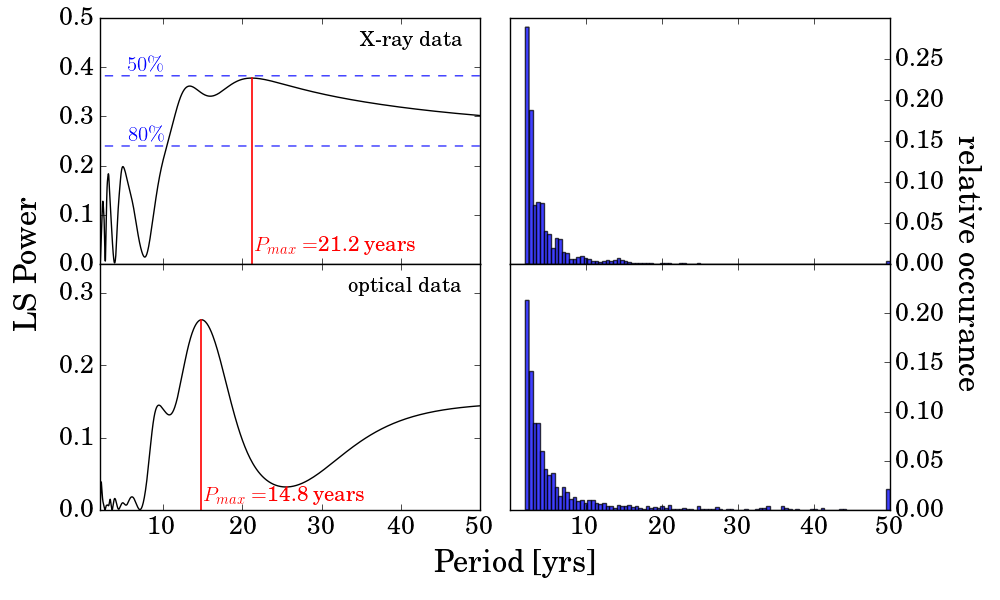}
\captionsetup{font=footnotesize}
\caption[width=0.5\textwidth]{ Lomb-Scargle periodograms and distribution of periods in the bootstrap approach of the X-ray and optical data.   Upper left panel: Lomb-Scargle periodogram of the X-ray data. The FAP levels of 50~$\%$ and 80~$\%$ resulting from the bootstrap analysis described in Section~\ref{pd} are marked as blue lines. Upper right panel: Distribution of maximum periods determined for all $10^4$ repetitions. Lower panels: Results of a similar analysis of the optical data. No FAP levels are displayed here, since none of the bootstrap repetitions yielded a higher Lomb-Scargle power than the unperturbed data.}
\label{periodogram}
\end{figure}

\begin{figure}[ht!]
\centering
\includegraphics[width=0.5\textwidth]{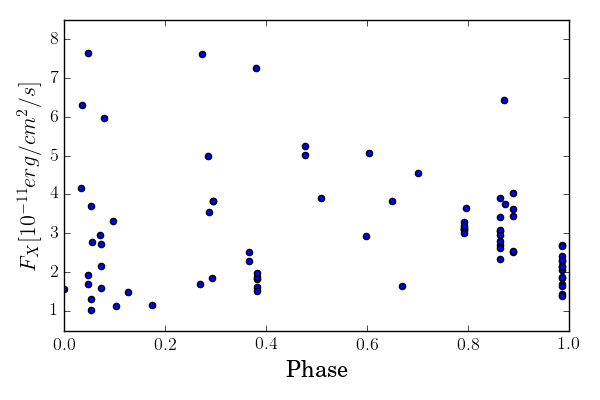}
\captionsetup{font=footnotesize}
\caption[width=0.5\textwidth]{Folded X-ray light curve under the assumption of the $21.2$~yr period determined in the LS analysis.}
\label{flc}
\end{figure}

In this Section we analyze possible periodicities in the X-ray data alone. The resulting X-ray light curve comprising of 80 observations is displayed in the top panel of Figure~\ref{xraylcunfolded}.\\ 
We excluded the SWIFT exposure ID 582894000 from further analysis on the grounds that the derived flux is two orders of magnitude larger compared to the rest of the data; this data set is the only SWIFT observation of the target connected to a TDRSS message, and therefore was likely triggered by an extreme flare of the system.\\
The remaining 79 data points were analyzed with the generalized Lomb-Scargle algorithm included in the PyAstronomy package with a normalization following \cite{2009A&A...496..577Z} while weighting each observation with the determined flux error; the upper left panel of Figure~\ref{periodogram} shows the resulting periodogram.\\
In order to assess the significance derived via the Lomb-Scargle periodogram, we ran a bootstrap approach with $10^4$ repetitions where, in every repetition, the fluxes were randomly drawn with replacement and distributed to the observation epochs.
The FAP is then the ratio between all repetitions with a higher LS power than that of the unperturbed analysis and the total number of repetitions. In the top left panel Figure~\ref{periodogram} we marked the FAP levels of 50~$\%$ and 80~$\%$ as dashed blue lines, and in the top right panel the maximum period distribution of all repetitions is displayed as a histogram.
While the periodogram exhibits a peak at $P=21.2$~yrs, the false alarm probability is higher than 50~$\%$, indicating that no significant period could be detected. A folded light curve assuming the putative period of $21.2$~yrs (Figure~\ref{flc}) confirms this finding, that is, at any given phase the entire range of fluxes is covered.\\
In order to check the influence of flaring events on the period analysis, we checked all available light curves for flares, and identified 33 observations without obvious large flaring events (see last column of Table~\ref{tab:data}). However, a period analysis of this reduced data set yielded no significant period, most likely due to insufficient temporal coverage. 

\subsubsection{Comparison to optical data}
\label{ai}
\begin{figure}[ht!]
\centering
\includegraphics[width=0.5\textwidth]{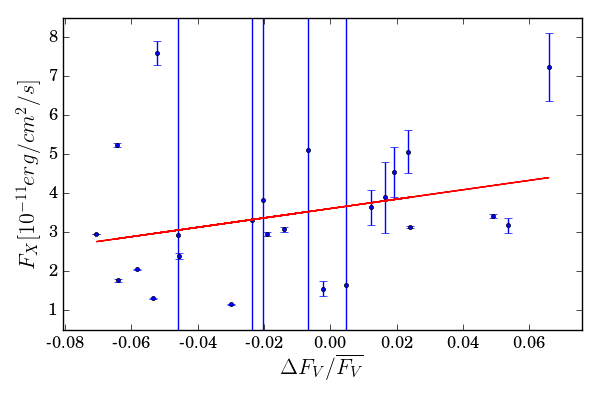}
\captionsetup{font=footnotesize}
\caption[width=0.5\textwidth]{Correlation between the optical and X-ray light curve. The best fit linear curve is marked with a red line.}
\label{cc}
\end{figure}

\begin{figure}[ht!]
\centering
\includegraphics[width=0.5\textwidth]{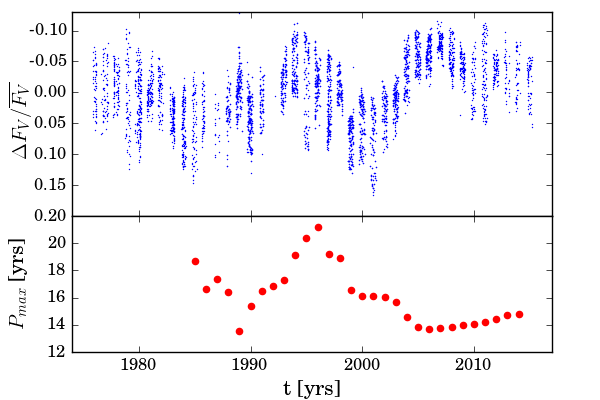}
\captionsetup{font=footnotesize}
\caption[width=0.5\textwidth]{Dependence of the maximum-likelihood period on temporal coverage of available data. Top: Photometric light curve in V-band. Bottom: Maximum periods derived via Lomb-Scargle analysis for data sets with increasing temporal coverage. The x-value of each point is the upper limit of the analyzed time window. The range of derived periods has a strong dependence on the time window.}
\label{rls}
\end{figure}

We performed a Lomb-Scargle analysis similar to the one described in Section~\ref{pd} on an optical data set comprising of archival data provided by \cite{2007ApJ...659L.157B}, \cite{2010A&A...521A..36M} and  \cite{2017ApJ...838..122J}. The resulting periodogram and distribution of periods in the bootstrap approach is displayed in the lower panels of Figure~\ref{periodogram}. Since none of the $10^4$ repetitions yielded a higher LS power than the maximum-likelihood period of $14.8$~yrs, no FAP could be determined, but rather an upper limit of $0.01\%$.\\
We tested the correlation between optical and X-ray data by binning each data set into bins of one year and then performed a cross-correlation between the two. The data yielded only a marginal correlation (see Figure~\ref{cc}) with a Pearson coefficient of $0.29$.
Computing a periodogram for different sets of archival optical data, starting with a window of ten years from the first data set and sequentially adding a year of data, we can show the effects of temporal coverage on derived periods. 
Figure~\ref{rls} shows the complete photometric light curve (top panel), as well as the period of maximum Lomb-Scargle power as a function of the upper limit of the time window. While the analysis of a sinusoidal data set with this method should result in near-constant values, we observe a large spread in the derived periods, which clearly depend on the time coverage of available data (as also evidenced by the previously mentioned disagreement between other authors). 

\section{Discussion and conclusions}
\label{discussion}
Our study of the longterm X-ray evolution of the active RS CVn binary HR~1099 has produced
the first X-ray light curve of the system spanning several decades. We stress that most data used for this work are available 
as standard pipeline products.\par 
We must nevertheless add some caveats:  The
X-ray instrumentation used substantially differs, some of the instruments are imaging,
others are non-imaging, some instruments offer high spectral resolution, others none and 
the covered spectral bands differ.  Some instruments cover only the energy
band above 2~keV (such as RXTE or MAXI); for those instruments HR~1099 appears as
a ''soft'' stellar source and the derived fluxes depend sensitively on the temperatures
used.  Other instruments, such as XMM-Newton cover the whole band pass, yet others, like the
ROSAT PSPC have little to no sensitivity at energies above 2~keV and therefore the
flux needs to be extrapolated, which again depends on the spectral model used.
We argue that our choices for the flux conversion are reasonable, however, we also 
admit that they are likely affected by systematic uncertainties that are beyond our control
such as the strength of the soft component during the RXTE observations and the strength of
the hard component during the ROSAT observations.\par 
Only a marginal correlation between the optical and X-ray data could be determined and we could not detect any significant periodicity in the X-ray data. Furthermore, the bootstrap method applied to estimate the FAP only accounts for white noise, and an approach including red noise would further decrease the significance of the detected peak \citep{2005A&A...431..391V,2018ApJS..236...16V}.
One possible explanation for the large FAP is the substantial scatter within the light curve, which is most likely dominated by stochastic X-ray flaring activity.\\
Another likely contributor is the shape of the activity modulation itself. Even when taking into account that the peak in the Lomb-Scargle periodograms of both X-ray and optical data is broad, indicating an error $\geq5$~yrs, the maximum-likelihood periods are only in partial agreement with the photometric cycles derived by other authors, namely \citet{2006A&A...455..595L} ($19.5\pm2$~yrs), \citet{2007ApJ...659L.157B} ($15-16$)~yrs and \citet{2010A&A...521A..36M} ($14.1\pm0.3$~yrs), as well as our own analysis of the complete data set ($14.8$~yrs). We thus conclude that the activity cycle of HR~1099 is more complicated than a simple sinusoidal, and we could only determine a characteristic timescale of 14.8~yrs based on the optical data.\par 
\cite{2006A&A...455..595L} claim a correlation of the magnetic cycle of HR~1099 with the orbital period variations, arguing that the orbital period variations display a period twice as long as that of the activity cycle, hinting at the Applegate mechanism. \cite{2010A&A...521A..36M} do, in fact, determine the period variations to be $36.3\pm1.9$ yr. 
However, in order to test this hypothesis, a continued long-term monitoring of HR~1099 and related objects using multiple activity indicators is required.
\\

\begin{acknowledgements} 
VP acknowledges funding through the DLR and DFG.
VP, DRGS, FHN and JAZF thank for funding through Fondecyt regular (project code 1161247). DRGS is grateful for funding through the ''Concurso Proyectos Internacionales de Investigaci\'on, Convocatoria 2015'' (project code PII20150171) and ALMA-Conicyt (project code 31160001). REM acknowledges funding via VRID-ENLACE 214.016.002-1.0 and VRID-ENLACE 218.016.004-1.0. DRGS, FHN and REM further thank for funding through the BASAL Centro de Astrof\'isica y Tecnolog\'ias Afines (CATA) PFB-06/2007. 
This research has made extensive use of data and/or software provided by the High Energy Astrophysics Science Archive Research Center (HEASARC), which is a service of the Astrophysics Science Division at NASA/GSFC and the High Energy Astrophysics Division of the Smithsonian Astrophysical Observatory. This research has made use of MAXI data provided by RIKEN, JAXA and the MAXI team. 
It is partly based on observations obtained with XMM-Newton, an ESA science mission with instruments and contributions directly funded by ESA Member States and NASA. 
The scientific results reported in this article are based in part on data obtained from the Chandra Data Archive. This work made use of data supplied by the UK Swift Science Data Centre at the
University of Leicester.
We would like to thank Steve Drake, Michael Corcoran and Bryan Irby (HEASARC) for help with the Einstein data set.
\end{acknowledgements}

\bibliography{eso.bib} %?? Get this done!
%\bibliography{report}   %>>>> bibliography data in report.bib
\bibliographystyle{aa}   %>>>> makes bibtex use spiebib.bst

\begin{appendix}
 \begin{table*}
\captionsetup{width=\linewidth}
\caption[width=\textwidth]{Data resulting from the X-ray analysis. Column one states the X-ray facility, and, in the case of multi-instrument missions, the instrument. The observing date is given as MJD and year in columns three and four. The mission-specific Observation ID and duration of the exposure are given in columns two and five.  The resulting X-ray flux and error in the energy range 2-10~keV are given in column six and seven.  The flux conversion factor (if data is derived from count rates) is given in column eight, the reduced $\chi^2$ for spectral fits in column nine and the flare flag in column ten indicates whether visual inspection of the light curve indicated a major flare during the observation, where 'y' denotes an obvious flaring event, 'n' stands for no large flare and '-' marks all observations for which a visual inspection was not possible.}
\label{tab:data}
\centering

\begin{tabular}{lccccccccc} 
\hline\hline 
Instrument & Obs. ID & MJD & year & t$_{stop}$-t$_{start}$  & $F_X$ & $\Delta F_X$ &conversion& ${\chi^2}_{red}$ & flare?\\
&&d&yyyy.yy&ks&$\frac{10^{-11}erg}{s\cdot cm^2}$&$\frac{10^{-11}erg}{s\cdot  cm^2}$&$\frac{10^{-11}erg}{s\cdot cm^2\cdot ct}$&&\\
\hline  \\
HEAO1&4U0336+01&43545.68&1978.10&4.6&1.56&0.23&-&1.10&n\\ 
EINSTEIN IPC&I03152&43900.23&1979.07&19.6&1.92&0.53&0.72&-&-\\ 
EINSTEIN SSS&shr1099a&43904.24&1979.08&0.7&7.65&1.08&-&1.75&y\\ 
EINSTEIN SSS&shr1099b&43905.23&1979.08&0.7&1.69&0.44&-&1.39&n\\ 
EINSTEIN IPC&I04496&44083.59&1979.57&1.6&2.97&0.82&0.72&-&-\\ 
EINSTEIN SSS&shr1099c&44095.70&1979.61&4.4&1.6&0.06&-&3.26&n\\ 
EINSTEIN SSS&shr1099d&44096.22&1979.61&5.3&2.15&0.25&-&1.24&n\\ 
EINSTEIN IPC&I02306&44098.93&1979.62&1.6&2.72&0.75&0.72&-&-\\ 
EINSTEIN IPC&I05455&44277.33&1980.10&4.1&3.33&0.92&0.72&-&-\\ 
EXOSAT ME&s10382&45624.92&1983.79&14.7&7.6&0.13&-&1.12&y\\ 
EXOSAT ME&s61328&46332.39&1985.73&32.8&2.28&0.09&-&1.33&n\\ 
EXOSAT ME&s61346&46332.72&1985.73&8.7&2.52&0.2&-&0.96&n\\ 
EXOSAT ME&s70753&46463.59&1986.09&42.5&1.97&0.08&-&1.69&n\\ 
EXOSAT ME&s70792&46464.08&1986.09&33.8&1.82&0.06&-&0.68&n\\ 
EXOSAT ME&s70825&46464.48&1986.09&29.9&1.63&0.08&-&1.31&n\\ 
EXOSAT ME&s70867&46464.85&1986.09&24.1&1.53&0.05&-&1.52&n\\ 
EXOSAT ME&s70893&46465.17&1986.09&22.6&1.89&0.1&-&0.93&n\\ 
Ginga LAC&g880125 232920&47186.98&1988.07&6.1&5.24&0.36&0.21&-&-\\ 
Ginga LAC&g880126 011132&47187.05&1988.07&233.1&5.01&0.35&0.21&-&-\\ 
RASS&RS931710N00&48102.26&1990.58&0.6&2.92&1.43&0.15&-&-\\ 
Ginga LAC&g910823 224507&48492.95&1991.65&1.1&3.82&0.38&0.21&-&-\\ 
ROSAT PSPC&rp200844n00&48648.44&1992.07&3&1.64&0.80&0.15&-&-\\ 
ASCA GIS2&22017000&49589.16&1994.65&41.9&3.09&0.04&-&1.18&y\\ 
ASCA GIS3&22017000&49589.16&1994.65&41.9&3.12&0.02&-&1.01&y\\ 
ASCA SIS0&22017000&49589.17&1994.65&37.6&3.14&0.07&-&1.08&y\\ 
ASCA SIS1&22017000&49589.17&1994.65&37.7&3.02&0.05&-&1.77&y\\ 
ASCA SIS0&22017000&49589.19&1994.65&24.7&3.29&0.04&-&1.23&y\\ 
ASCA SIS1&22017000&49589.19&1994.65&24.8&3.2&0.06&-&1.33&-\\ 
ROSAT HRI&RH202286N00&50126.43&1996.12&5.4&2.34&0.64&1.13&-&-\\ 
ROSAT HRI&RH202288N00&50126.83&1996.12&5.2&2.69&0.73&1.13&-&-\\ 
ROSAT HRI&RH202295N00&50127.37&1996.12&4.2&3.06&0.83&1.13&-&-\\ 
ROSAT HRI&RH202293N00&50127.76&1996.12&4&3.41&0.93&1.13&-&-\\ 
ROSAT HRI&RH202291N00&50128.31&1996.12&3.9&2.81&0.76&1.13&-&-\\ 
ROSAT HRI&RH202290N00&50128.72&1996.12&3.7&3.92&1.10&1.13&-&-\\ 
ROSAT HRI&RH202289N00&50129.31&1996.13&3.2&3.09&0.84&1.13&-&-\\ 
ROSAT HRI&RH202287N00&50129.65&1996.13&2.2&2.63&0.72&1.13&-&-\\ 
ROSAT HRI&RH202292N00&50130.58&1996.13&1.8&2.96&0.80&1.13&-&-\\ 
RXTE PCA&10005010100&50328.81&1996.67&9.6&2.54&0.11&-&1.29&y\\ 
RXTE PCA&10005010101&50329.75&1996.67&8.9&2.52&0.12&-&1.19&n\\ 
RXTE PCA&10005010102&50330.76&1996.68&8.5&3.44&0.14&-&2.67&y\\ 
RXTE PCA&10005010103&50331.95&1996.68&3.3&4.03&0.19&-&0.83&n\\ 
RXTE PCA&10005010104&50332.01&1996.68&7.8&3.63&0.14&-&1.53&n\\ 
BeppoSAX MECS&1045501&51062.92&1998.68&130.9&1.71&0.04&-&1.15&n\\ 
BeppoSAX LECS&1045501&51062.92&1998.68&62.9&1.45&0.04&-&1.77&n\\ 
RXTE PCA&30003010102&51063.64&1998.68&1.6&1.39&0.08&-&1.13&n\\ 
RXTE PCA&30003010108&51063.77&1998.68&9.9&2.42&0.13&-&0.71&n\\ 
RXTE PCA&30003010100&51064.71&1998.69&15.9&2.31&0.1&-&1.83&n\\ 
RXTE PCA&30003010100&51065.04&1998.69&1.2&1.86&0.29&-&0.84&n\\ 
RXTE PCA&30003010107&51065.13&1998.69&10&2.15&0.08&-&0.76&n\\ 
RXTE PCA&30003010101&51065.66&1998.69&14.6&2.69&0.19&-&2.13&n\\ 
RXTE PCA&30003010101&51065.97&1998.69&4.4&1.88&9.54&-&1.10&n\\ 
RXTE PCA&30003010104&51066.12&1998.69&3.2&2.06&0.14&-&0.75&n\\ 
RXTE PCA&30003010103&51066.64&1998.69&16&2.67&0.12&-&1.86&n\\ 
\end{tabular}
\end{table*}

 \begin{table*}
\captionsetup{width=\linewidth}
\caption*{Table \ref{tab:data} continued...}
\centering

\begin{tabular}{lccccccccc} 
\hline\hline 
Instrument & Obs. ID & MJD & year & t$_{stop}$-t$_{start}$  & $F_X$ & $\Delta F_X$ &conversion & ${\chi^2}_{red}$ & flare?\\
&&d&yyyy.yy&ks&$\frac{10^{-11}erg}{s\cdot cm^2}$&$\frac{10^{-11}erg}{s\cdot  cm^2}$&$\frac{10^{-11}erg}{s\cdot cm^2\cdot ct}$&&\\
\hline \\
RXTE PCA&30003010103&51066.97&1998.69&2.4&1.65&0.24&-&0.75&n\\ 
RXTE PCA&30003010106&51067.00&1998.69&0.8&2.14&0.38&-&0.96&n\\  
RXTE PCA&30003010105&51067.04&1998.69&7&2.29&0.14&-&0.78&n\\ 
Chandra ACIS-S HETG&62538&51435.95&1999.79&95&4.17&0.05&-&1.00&y\\
Chandra ACIS-S HETG&1252&51438.53&1999.80&15&6.31&0.14&-&1.00&n\\
XMM-Newton PN&0116200701&51575.39&2000.08&15.1&1.04&0.02&-&1.01&n\\ 
XMM-Newton PN&0116340601&51577.52&2000.09&28.6&1.32&0.02&-&1.17&n\\ 
XMM-Newton PN&0116890901&51582.83&2000.10&40&3.7&0.05&-&1.10&y\\ 
XMM-Newton PN&0117890901&51592.59&2000.13&57.7&2.77&0.02&-&1.00&y\\ 
XMM-Newton PN&0129350201&51784.39&2000.66&31.3&5.96&0.06&-&1.10&y\\ 
XMM-Newton PN&0134540101&51962.75&2001.15&46.3&1.13&0.02&-&1.21&y\\ 
XMM-Newton PN&0134540401&52139.16&2001.63&26.4&1.48&0.02&-&1.05&y\\ 
XMM-Newton PN&0134540601&52508.82&2002.64&35.7&1.15&0.01&-&1.31&y\\ 
XMM-Newton MOS1&0134540801&53230.76&2004.62&51.9&1.71&0.05&-&0.86&y\\ 
SWIFT XRT&50850001&53352.40&2004.95&0.4&5&0.74&-&1.24&y\\ 
SWIFTXRT&50850002&53359.23&2004.97&5.3&3.55&0.53&-&1.13&y\\ 
SWIFT XRT&50850008&53419.04&2005.13&9.1&1.85&0.27&-&1.07&y\\ 
SWIFT XRT&50850009&53423.05&2005.14&13.9&3.84&0.57&-&0.91&y\\ 
SWIFT XRT&30837001&54069.87&2006.91&1.9&7.24&1.08&-&1.10&y\\ 
MAXI GSC+SSC&---&55058.52&2009.62&1463&3.9&0.64&-&1.05&y\\ 
MAXI GSC+SSC&---&55788.52&2011.62&1122&5.07&0.52&-&1.06&y\\ 
MAXI GSC+SSC&---&56519.52&2013.62&919.4&4.55&0.67&-&1.05&y\\ 
SWIFT XRT&582894000&56661.66&2014.01&1&319.22&47.46&-&1.03&y\\ 
MAXI GSC+SSC&---&57249.52&2015.62&933.3&3.64&0.45&-&1.13&y\\ 
SWIFT XRT&88061001&57833.06&2017.22&3&6.44&0.96&-&0.63&n\\ 
SWIFT XRT&88061002&57835.00&2017.22&1.6&3.75&0.56&-&1.17&n\\ 
\end{tabular}
\end{table*}

\end{appendix}

\end{document}